# Theory of topological insulator waveguides: polarization control and the enhancement of the magento-electric effect

J. A. Crosse*

Department of Electrical and Computer Engineering, National University of Singapore, 4 Engineering Drive 3, Singapore 117583.

*e-mail: alexcrosse@gmail.com*

**Topological insulators[1,2] subject to a time-symmetry-breaking perturbation are predicted to display a magneto-electric effect that causes the electric and magnetic induction fields to mix at the material's surface[3,4]. This effect induces polarization rotations of between ≈1-10 mrad per interface in incident plane-polarized light normal to a multilayered structure[5-8]. Here we show, theoretically and numerically, that, using a waveguide geometry with a topological insulator guide layer and dielectric cladding, it is possible to achieve rotations of between ≈100-1000 mrad and generate an elliptical polarization with only a three-layered structure. Both the rotation angle and ellipticity are dependent on the permittivity contrast of the guide and cladding layers and the strength of the time-symmetry-breaking perturbation. This geometry is beneficial, not only as a way to enhance the magneto-electric effect, rendering it easier to observe, but also as a method for controlling the polarization of light in the next generation of photonic devices.**

Topological insulators are time-reversal symmetric materials that display non-trivial topological order, which results in an insulating bulk and protected conducting edge states[9]. Such behavior has been observed in Group V/VI alloys that have strong enough spin-orbit coupling to induce band inversion, e.g. $Bi_{1-x}Sb_x$[10,11], $Bi_2Se_3$, $Bi_2Se_3$ and $Sb_2Te_3$[12-14]. Topological insulators can also be used to realize an axionic material[15]. To achieve this one needs to apply a time-symmetry-breaking perturbation of sufficient size to the material's surface to open a gap. Such a perturbation can be achieved by either introducing ferromagnetic dopants[4] – 12% dopants opens a 50 meV gap[16] – or by applying an external magnetic field[5]. The resulting Hall effect leads to modified constitutive relations for the material

$$D(r,\omega) = \varepsilon(r,\omega)E(r,\omega) + \frac{\alpha}{\pi}\Theta(r,\omega)B(r,\omega),$$

(1)

$$H(r,\omega) = \frac{1}{\mu(r,\omega)}B(r,\omega) + \frac{\alpha}{\pi}\Theta(r,\omega)E(r,\omega),$$

(2)

where $\varepsilon(r,\omega)$ and $\mu(r,\omega)$ are the usual permittivity and permeability and $\Theta(r,\omega)$ is the axion coupling which takes odd values of $\pi$ in an time-symmetry-broken topological insulator and even values of $\pi$ in a conventional magneto-dielectric. Here, α, is the fine structure constant. For convenience we will work in natural units with c = $\varepsilon_0$ = $\mu_0$ = 1. One can see that the axion coupling causes the $E(r,\omega)$ and $B(r,\omega)$ fields to mix. Although Maxwell's equations are unchanged by the axion coupling, the wave equation gains a term proportional to $\nabla\Theta(r,\omega) \times E(r,\omega)$[8,15,17]. Thus, gradients in the axion coupling cause the electric field to rotate in the plane perpendicular to that gradient. Hence, plane polarized light will undergo a Kerr (reflection) and Faraday (transmission) rotations in regions of spatially varying axion coupling.

Here, we study a slab waveguide geometry, layered in the y-direction with optical modes propagating in the x-direction, consisting of a time-symmetry-broken topological insulator guide layer and a conventional magneto-dielectric cladding. Both the guide layer and the cladding are considered to be homogeneous. Thus, the only variation in the axion coupling occurs at the interfaces between the guide layer and the cladding. Such a waveguide supports four types of mode, transverse electric odd and even modes (TE) where the electric field, $E(r,\omega)$, is polarized in the z-direction and transverse magnetic odd and even modes (TM) where the magnetic field, $H(r,\omega)$, is polarized in the z-direction. As the effect of the axion coupling is to rotate the polarization of the propagating field, one would expect the TE and TM polarizations to mix. Thus, one should observe a rotation in the polarization of plane-polarized light as it propagates down the waveguide. Furthermore, as the propagation constants of the TE and TM modes are different in a waveguide, one should also see the light become elliptically polarized, with the ellipticity varying as the light propagates down the waveguide. From a geometric optics perspective, one can view the effect as a series of successive Kerr rotations from the multiple reflections at the waveguide interfaces. In this geometry, there are two possible orientations for the axion coupling. The first is anti-parallel orientation (see Fig 1a), where the change in the axion coupling is the same on each side of the waveguide. In this case the interaction rotates light in the same direction with respect to the direction of propagation.

Hence, magnetic fields of the same sign (and electric fields of opposite sign) are generated at opposing interfaces. Thus, even modes will couple to even modes and odd modes will couple to odd modes (see Fig. 1b). The second case is the parallel case (see Fig 1c), where the change in the axion coupling is equal and opposite on each side of the waveguide. In this case the interaction rotates the light in opposite directions with respect to the direction of propagation. Hence, magnetic fields of opposite sign (and electric fields of the same sign) are generated at opposing interfaces. Thus, even modes will couple to odd modes visa versa (see Fig. 1d).

To quantify this effect we use the modified wave equation to develop a coupled mode theory for the system (see supplementary information). Owing to the relative sizes of α and ε, we assume that the axion coupling term only marginally changes the modes of the waveguide and, hence, can treat it as a small perturbation to the usual magneto-dielectric waveguide modes. Using these modes as a basis, one finds that the mode amplitudes, $\tilde{A}_\nu(x)$, evolve as

$$\frac{\partial^2}{\partial x^2}\tilde{A}_\nu(x) + \beta_\nu^2 \tilde{A}_\nu(x) + \sum_\mu \kappa_{\nu\mu}\tilde{A}_\mu(x) = 0,$$

(3)

with the coupling constants, $\kappa_{\nu\mu}$, give by

$$\kappa_{\nu\mu} = i\omega\frac{\alpha}{\pi}\int_{-\infty}^{\infty} dy \left\{ [\boldsymbol{\nabla}\Theta(y,\omega) \times \boldsymbol{\mathcal{E}}_\mu(y)] \times \boldsymbol{\mathcal{H}}_\nu^*(y) + [\boldsymbol{\nabla}\Theta(y,\omega) \times \boldsymbol{\mathcal{H}}_\mu(y)] \times \boldsymbol{\mathcal{E}}_\nu^*(y) \right\} \cdot \hat{x}.$$

(4)

where $\boldsymbol{\mathcal{E}}_\nu(y)$ and $\boldsymbol{\mathcal{H}}_\nu(y)$ are the usual electric and magnetic field profiles for a magneto-dielectric waveguide and $\beta_\nu$ are the corresponding propagation constants. Here, the indices μ and ν indicate the TE, TM, odd and even modes. Considering only the lowest four modes, the interaction only couples pairs of modes. Hence, Eq. 3 breaks down into two independent 2x2 matrix equations that can be solved analytically. As an example we consider the anti-parallel configuration with the input light exciting the TE even mode only. Evaluation of the coupling constants shows that, in this situation, the interaction only couples the input light to the TM even mode. Solving the resulting system of equations results shows that the $E_y(x)$ and $E_z(x)$ components of the field evolve as the sum of two waves of different frequencies. By analyzing the beat properties between the two waves one can find the coupling length, $l_c$, and maximum transferred power, $P_c$. In weak coupling limit, $\kappa_{TE,TM}, \kappa_{TM,TE} \ll \beta_{TE}^2, \beta_{TM}^2$, and which is always valid for the parameters considered here, one finds

$$l_c \approx \frac{\pi\sqrt{2(\beta_{TE}^2 + \beta_{TM}^2)}}{\lambda_0},$$

(5)

and

$$P_c \approx 4\left|\frac{\kappa_{TM,TE}}{\lambda_0}\right|^2.$$

(6)

with $\lambda_0 = \sqrt{(\beta_{TE}^2 + \beta_{TM}^2)^2 + 4\kappa_{TE,TM}\kappa_{TM,TE}}$. From the fields component one can also find the maximum polarization rotation, which occurs at the coupling length, $l_c$, and maximum ellipticity, which occurs at $\frac{l_c}{2}$. In the weak coupling limit, these read

$$\psi_{max} \approx \frac{1}{2}\tan^{-1}\left(\frac{2\sqrt{P_c(1-P_c)}}{1-2P_c}\right),$$

(7)

$$\chi_{max} \approx \frac{1}{2}\sin^{-1}\left(\frac{\sqrt{P_c}(2-P_c)}{2\sqrt{1-P_c}}\right).$$

(8)

In order to validate the theoretical model we compared the results to FDTD[18] electromagnetic simulation data. We consider a $Bi_2Se_3$ guide layer, with $\varepsilon_g = 16$ (valid at high frequencies[19]) and a silicon cladding, with $\varepsilon_c = 12$ and insert light of wavelength ≈3.3 times the waveguide width. For both the guide layer and the cladding we take µ = 1. At this frequency there are four guided modes - the lowest order TE and TM, odd and even modes. Figure 2a shows simulated data of the elliptical polarization of the light at different points along the waveguide. One observes the expected changes in rotation and ellipticity as the light propagates. Figure 2b shows the elliptical polarization of the light at ≈22 wavelengths along the waveguide for both theoretical calculation and simulated data. One observes good agreement between the two. Figure 2c shows the variation in the polarization angle and ellipticity along the length of the waveguide. The polarization and ellipticity are seen to oscillate with a wavelength of, $4l_c$, with the maximum rotation occurring at the coupling length and maximum ellipticity occurring at half the coupling length. The phenomenological shift of 4 mrad that occurs in the simulated data is a result of the source partially exciting the TM even mode hence the initial polarization is 4 mrad away from the z-axis. Good agreement is seen between the numerical calculation and simulated data.

Now we consider how to maximize the polarization rotation and ellipticity of the input light. The coupling constants are directly proportional to the change in the axion coupling at the waveguide edges. Increasing, this change by increasing the strength of the time-symmetry-breaking perturbation, would lead to a larger effect. Since an external magnetic field can be used to provide the perturbation, magnetically controlled rotations are possible with this type of device. The other way to change the coupling constants would be to change the waveguide mode profiles. These profiles are directly related to the dielectric properties of the waveguide guide layer and cladding. Figure 3a shows the maximum polarization rotation and ellipticity for different permittivity contrasts, $\varepsilon_g - \varepsilon_c$. One sees that as the contrast reduces the maximum polarization rotation and ellipticity increase and for contrasts of ≈0.5 one sees rotations of ≈100 mrads (≈6°), which is two orders of magnitude larger than one sees for transmission or reflection from a single interface. A large coupling length accompanies the large rotation. Hence, to observe such a rotation very long waveguides need to be fabricated (≈1000 wavelengths). This effect is shown in Fig. 3b. The increase in rotation is due to the improved phase matching between the TE even and TM even modes. As the permittivity contrast is reduced the phase mismatch, $\beta_{TE} - \beta_{TM}$, reduces and hence there is better transfer of energy between the modes. However, reducing the permittivity contrast reduces the coupling constant. This effect is smaller than the change in the phase mismatch (a factor of ≈2 over the parameters studied compared to a order of magnitude for the phase mismatch) but contributes to the increase in coupling length.

Topological insulator waveguides offer an easily fabricatable structure that can increase the rotation of plane-polarized light, a key signature in the detection of the magneto-electric effect, by two orders of magnitude compared to other three-layered structures. Furthermore, the structure dependent, magnetically controllable coupling between the waveguide modes has the potential to offer robust polarization control in the next generation of photonic devices.

**Methods**

Using the constitutive equations to eliminate the magnetic induction, $\boldsymbol{B}(\boldsymbol{r},t)$, and displacement, $\boldsymbol{D}(\boldsymbol{r},t)$, fields from the time dependent Maxwell's equations and neglecting terms of order $O(\alpha^2)$ leads to two coupled equations for the electric, $\boldsymbol{E}(\boldsymbol{r},t)$, and magnetic, $\boldsymbol{H}(\boldsymbol{r},t)$, fields

$$\varepsilon(\boldsymbol{r})\frac{\partial}{\partial t}\boldsymbol{E}(\boldsymbol{r},t) = \nabla \times \boldsymbol{H}(\boldsymbol{r},t) + \frac{\alpha}{\pi}\Theta(\boldsymbol{r})\nabla \times \boldsymbol{E}(\boldsymbol{r},t),$$

(9)

$$\varepsilon(\boldsymbol{r})\frac{\partial}{\partial t}\boldsymbol{H}(\boldsymbol{r},t) = -\varepsilon(\boldsymbol{r})\nabla \times \boldsymbol{H}(\boldsymbol{r},t) - \frac{\alpha}{\pi}\Theta(\boldsymbol{r})\nabla \times \boldsymbol{E}(\boldsymbol{r},t).$$

(10)

Equations (9) and (10), subject to a unit plane wave source, $E_z = -H_y = e^{-i\omega t}$, were solved on a staggered Yee lattice using standard electromagnetic FDTD methods[18].

**Acknowledgements** We acknowledge supports from the Singapore National Research Foundation under NRF Grant No. NRF-NRFF2011-07. The author would like to thank S. Y. Buhmann and M. Tsang for useful discussions.

**Author contributions** The author conceived the idea, developed the theory, carried out the numerical calculation and wrote the paper.

**Additional information** The author declares no competing financial interests. Supplementary information accompanies this paper. Correspondence and requests for materials should be addressed to J.A.C.

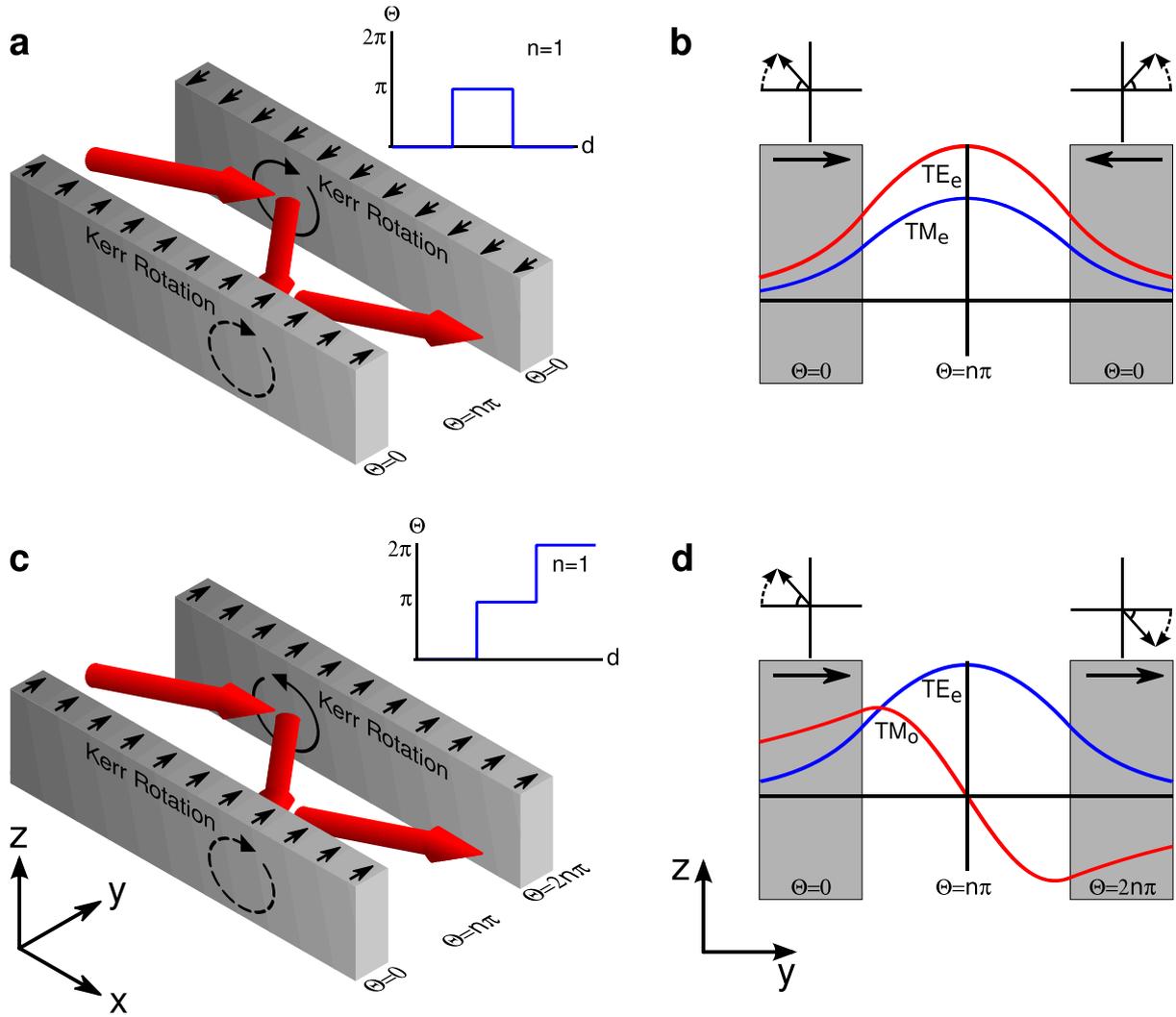

**Figure 1: Schematic diagrams of time-symmetry-broken topological insulator waveguides. (a)** The anti-parallel configuration. The time-symmetry breaking perturbations at each interface of the waveguide are in opposite directions. The inset shows the axion coupling profile for this configuration for n=1. **(b)** As the time-symmetry breaking perturbations at each interface of the waveguide are in opposite directions, the rotations of the magnetic fields at each interface are in opposite directions and, hence, the TE and TM even modes couple. **(c)** The parallel configuration. The time-symmetry breaking perturbations at each interface of the waveguide are in the same directions. The inset shows the axion coupling profile for this configuration for n=1. **(d)** As the time-symmetry breaking perturbations at each interface of the waveguide are in the same direction, the rotations of the magnetic

fields at each interface are in the same direction and, hence, the TE even and TM odd modes couple.

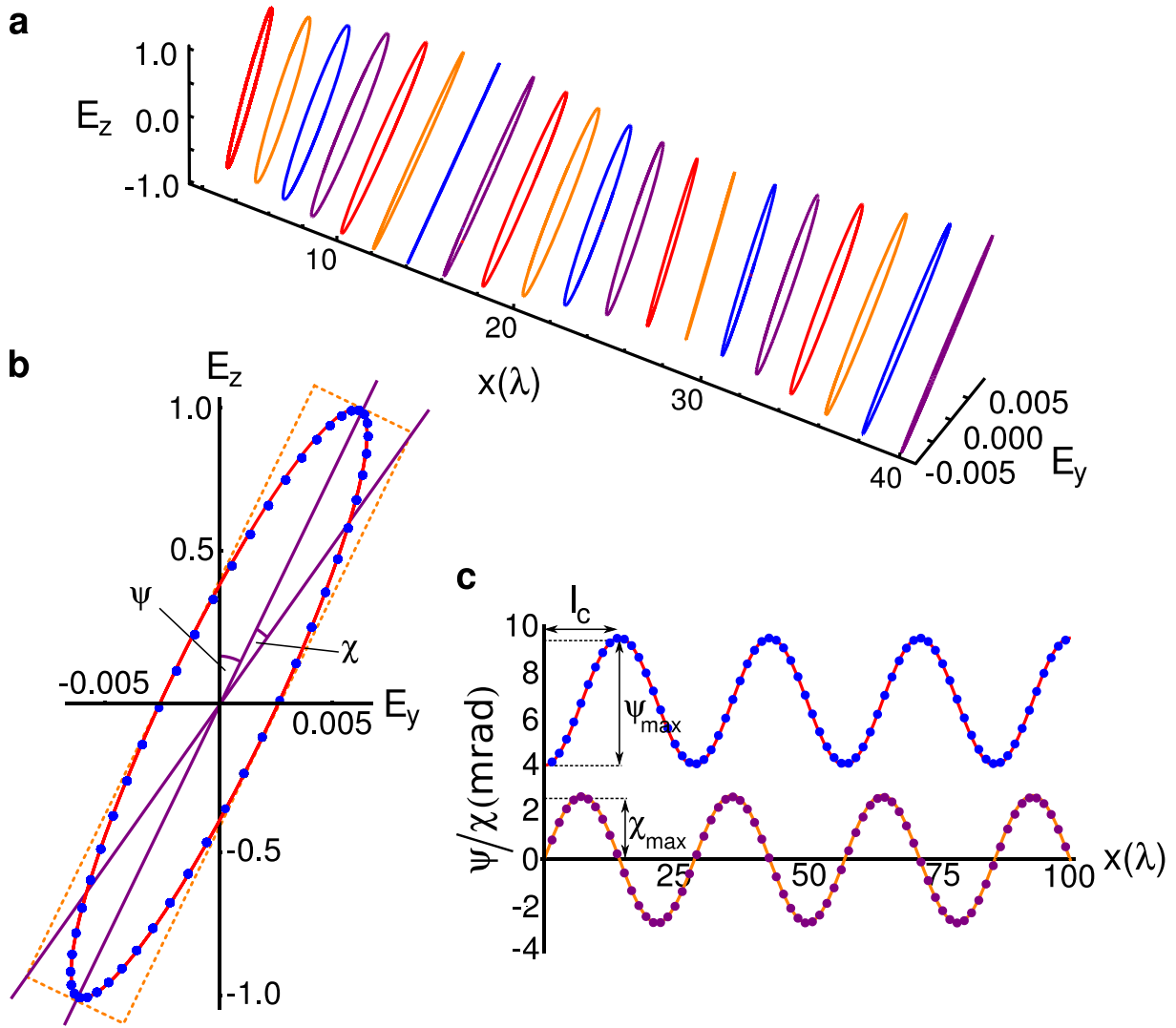

**Figure 2: Variation of the polarization angle and ellipticity with distance in a time-symmetry-broken topological insulator waveguide.** (a) Simulated data of the polarization and ellipticity of plane polarized light as a function of distance along the waveguide. (b) The ellipticity and polarization angle of light at ≈22 wavelengths along the waveguide. The red line shows the theoretically calculated values and the solid blue circles show the simulated data. The angles marked $\psi$ and $\chi$ are the polarization angle and ellipticity, respectively. (c) The polarization angle and

ellipticity as a function of distance along the waveguide. The red and orange lines are the theoretically calculated values and the solid blue and purple circles are the simulated data. The maximum polarization angle occurs at the coupling length, $l_c$, and the largest ellipticity occurs at half the coupling length, $\frac{l_c}{2}$. The 4 mrad offset in the polarization angle is due to partial excitation of the TM even mode.

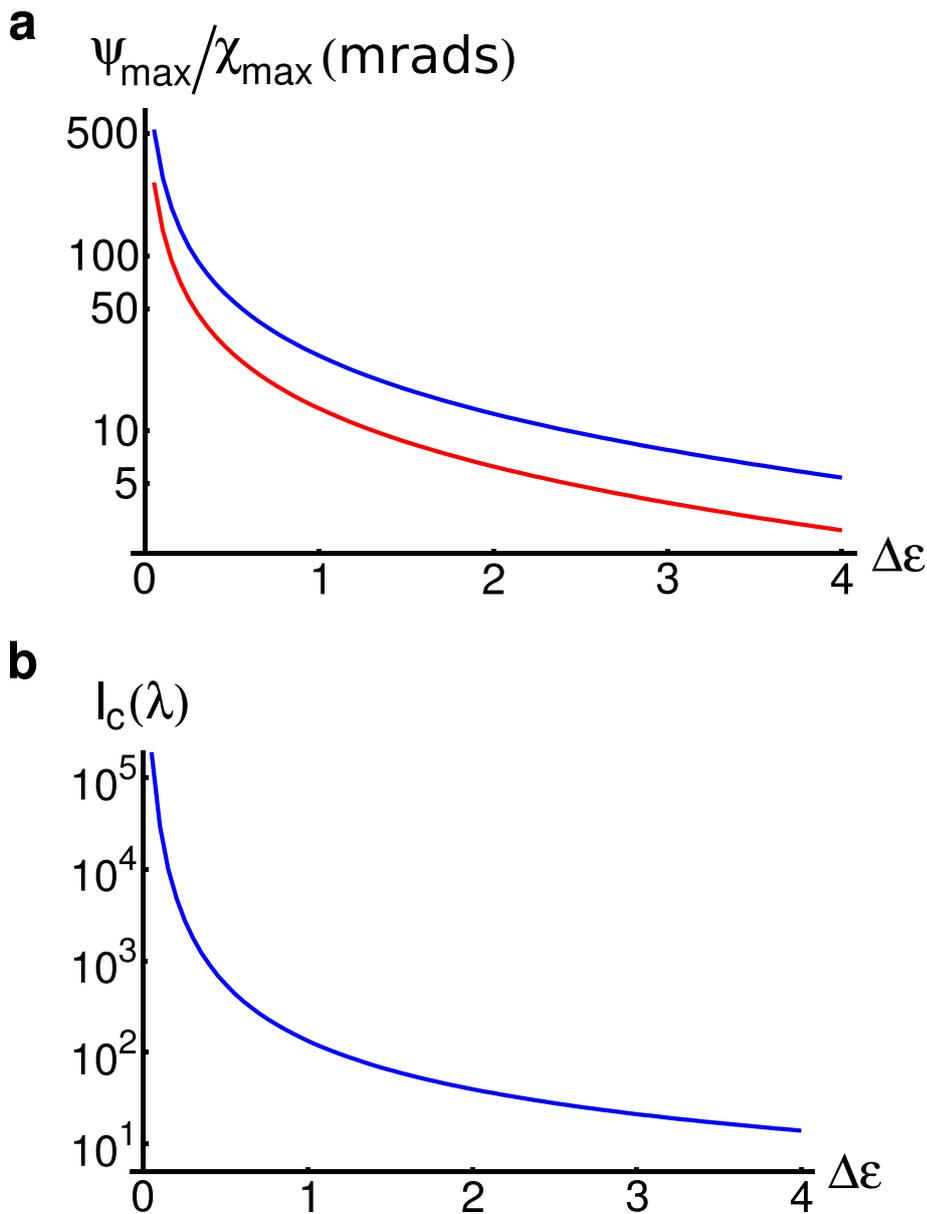

**Figure 3: Variation of the polarization angle, ellipticity and coupling length with permittivity contrast in a time-symmetry-broken topological insulator**

**waveguide. (a)** Log-linear plot of theoretically calculated values of the variation in the maximum polarization angle, $\psi_{max}$, (blue line) and ellipticity, $\chi_{max}$, (red line) as a function permittivity contrast. **(b)** Log-linear plot of theoretically calculated values of the variation in the coupling length, $l_c$, as a function of permittivity contrast.

# SUPPLEMENTARY INFORMATION for Theory of topological insulator waveguides: polarization control and the enhancement of the magento-electric effect

J. A. Crosse*,

Department of Electrical and Computer Engineering, National University of Singapore, 4 Engineering Drive 3, Singapore 117583.

*e-mail: alexcrosse@gmail.com*

Consider a slab waveguide consisting of three layers, layered in the y-direction, with permittivity profile, $\varepsilon(y,\omega)$, permeability profile, $\mu(y,\omega)$, and axion coupling profile, $\Theta(y,\omega)$. We arrange our coordinate system such that the guided modes propagate in the x-direction. From the source free Maxwell Equations and constitutive relations for a time-symmetry broken topological insulator one can show that, in the presence of a non-vanishing axion coupling, the electric, $\mathbf{E}(\mathbf{r},\omega)$, and magnetic, $\mathbf{H}(\mathbf{r},\omega)$, fields obey the wave equations

$$\nabla^2 \mathbf{E}(\mathbf{r},\omega) - \omega^2 \varepsilon(y,\omega)\mathbf{E}(\mathbf{r},\omega) - i\omega\frac{\alpha}{\pi}[\nabla\Theta(y,\omega) \times \mathbf{E}(\mathbf{r},\omega)] = \mathbf{0},$$

(1a)

$$\nabla^2 \mathbf{H}(\mathbf{r},\omega) - \omega^2 \varepsilon(y,\omega)\mathbf{H}(\mathbf{r},\omega) + i\omega\frac{\alpha}{\pi}[\nabla\Theta(y,\omega) \times \mathbf{H}(\mathbf{r},\omega)] = \mathbf{0},$$

(1b)

respectively. Here, for convenience, we work in natural units where $\mu_0 = \varepsilon_0 = c = 1$ and we have assumed that magnetic effects are negligible and, hence, $\mu(y,\omega) = 1$. Furthermore, as α is of the order $10^{-2}$ compared to ε, terms of order $\alpha^2$ are also negligible.

Owing to the relative sizes of α and ε, we assume that the axion coupling only marginally changes the modes of the waveguide and hence we can treat the last terms in equation Eqs. (1a) and (1b) as a small perturbation to the usual waveguide modes of a traditional magneto-dielectric waveguide. Hence we can expand the electric and magnetic field as

$$\mathbf{E}(\mathbf{r},\omega) = \sum_{\mu} A_{\mu}(x)\boldsymbol{\mathcal{E}}_{\mu}(y)e^{i\beta_{\mu}x},$$

(2a)

$$\mathbf{H}(\mathbf{r},\omega) = \sum_{\mu} A_{\mu}(x)\boldsymbol{\mathcal{H}}_{\mu}(y)e^{i\beta_{\mu}x},$$

(2b)

where $\boldsymbol{\mathcal{E}}_{\mu}(y)$ and $\boldsymbol{\mathcal{H}}_{\mu}(y)$ are the usual electric and magnetic field profiles of mode μ for a magneto-dielectric waveguide and $\beta_{\mu}$ and $A_{\mu}(x)$ are the corresponding propagation constant and amplitude. Substituting the expansions in Eqs. (2a) and (2b) into Eqs. (1a) and (1b) and noting that, by definition, the dielectric waveguide modes obey

$$\frac{\partial^2}{\partial y^2}\boldsymbol{\mathcal{E}}_{\mu}(y) + [\omega^2\varepsilon(y,\omega) - \beta_{\mu}^2]\boldsymbol{\mathcal{E}}_{\mu}(y) = \mathbf{0},$$

(3a)

$$\frac{\partial^2}{\partial y^2}\boldsymbol{\mathcal{H}}_{\mu}(y) + [\omega^2\varepsilon(y,\omega) - \beta_{\mu}^2]\boldsymbol{\mathcal{H}}_{\mu}(y) = \mathbf{0},$$

(3b)

one finds that the wave equations become

$$\sum_{\mu}\left[\frac{\partial^2}{\partial x^2}\tilde{A}_{\mu}(x) + \beta_{\mu}^2\tilde{A}_{\mu}(x)\right]\boldsymbol{\mathcal{E}}_{\mu}(y) + i\omega\frac{\alpha}{\pi}\sum_{\mu}[\nabla\Theta(y,\omega) \times \boldsymbol{\mathcal{E}}_{\mu}(y)]\tilde{A}_{\mu}(x) = \mathbf{0},$$

(4a)

$$\sum_{\mu}\left[\frac{\partial^2}{\partial x^2}\tilde{A}_{\mu}(x) + \beta_{\mu}^2\tilde{A}_{\mu}(x)\right]\boldsymbol{\mathcal{H}}_{\mu}(y) - i\omega\frac{\alpha}{\pi}\sum_{\mu}[\nabla\Theta(y,\omega) \times \boldsymbol{\mathcal{H}}_{\mu}(y)]\tilde{A}_{\mu}(x) = \mathbf{0},$$

(4b)

where $\tilde{A}_{\mu}(x) = e^{i\beta_{\mu}x}A_{\mu}(x)$. Using the orthogonality relation

$$\int_{-\infty}^{\infty}dy\,[\boldsymbol{\mathcal{E}}_{\mu}(y) \times \boldsymbol{\mathcal{H}}_{\nu}^{*}(y) + \boldsymbol{\mathcal{E}}_{\nu}^{*}(y) \times \boldsymbol{\mathcal{H}}_{\mu}(y)] \cdot \hat{x} = \delta_{\mu\nu},$$

(5)

where $\delta_{\mu\nu}$ is the Kronecker delta, one can show that the coupled mode equations for a specific mode, $A_{\mu}$, read

$$\frac{\partial^2}{\partial x^2}\tilde{A}_{\nu}(x) + \beta_{\nu}^2\tilde{A}_{\nu}(x) + \sum_{\mu}\kappa_{\nu\mu}\tilde{A}_{\mu}(x) = 0,$$

(6)

with the coupling constant $\kappa_{\nu\mu}$ is defined as

$$\kappa_{\nu\mu} = i\omega\frac{\alpha}{\pi}\int_{-\infty}^{\infty} dy\left\{[\nabla\Theta(y,\omega)\times\mathcal{E}_{\mu}(y)]\times\mathcal{H}_{\nu}^{*}(y) + [\nabla\Theta(y,\omega)\times\mathcal{H}_{\mu}(y)]\times\mathcal{E}_{\nu}^{*}(y)\right\}\cdot\hat{x}.$$

(7)

For a slab waveguide with a guide layer of width, d, centered at $y = 0$ and constant axion couplings $\Theta_{+}$, $\Theta_{-}$ and $\Theta_{g}$, in the upper, lower and guide layers respectively one can write the overall axion coupling as

$$\Theta(y,\omega) = \Theta_{-}\text{H}\left(-y-\frac{d}{2}\right) + \Theta_{g}\text{H}\left(y+\frac{d}{2}\right)\text{H}\left(-y+\frac{d}{2}\right) + \Theta_{+}\text{H}\left(y-\frac{d}{2}\right),$$

(8)

where H($y$) is the Heaviside step function. The derivative of the axion coupling profile reads

$$\nabla\Theta(y,\omega) = (\Theta_{g}-\Theta_{-})\delta\left(-y-\frac{d}{2}\right) - (\Theta_{g}-\Theta_{+})\delta\left(y-\frac{d}{2}\right).$$

(9)

Thus one sees that the coupling term vanishes everywhere except at the interface of the guide layer and the cladding. Substituting, Eq. (9) into the definition of the coupling constant in Eq. (7) and evaluating the cross products leads to

$$\kappa_{\nu\mu} = i\beta_{\nu}\frac{\alpha}{\pi}(\Theta_{g}-\Theta_{+})\left\{\frac{\mathcal{H}_{\nu,z}^{*}\left(\frac{d}{2}\right)\mathcal{H}_{\mu,x}\left(\frac{d}{2}\right)}{\varepsilon_{g}} - \mathcal{E}_{\nu,z}^{*}\left(\frac{d}{2}\right)\mathcal{E}_{\mu,x}\left(\frac{d}{2}\right)\right\}$$

$$-i\beta_{\nu}\frac{\alpha}{\pi}(\Theta_{g}-\Theta_{-})\left\{\frac{\mathcal{H}_{\nu,z}^{*}\left(-\frac{d}{2}\right)\mathcal{H}_{\mu,x}\left(-\frac{d}{2}\right)}{\varepsilon_{g}} - \mathcal{E}_{\nu,z}^{*}\left(-\frac{d}{2}\right)\mathcal{E}_{\mu,x}\left(-\frac{d}{2}\right)\right\},$$

(10)

which shows that the effect of the axion coupling is to rotate the transverse electric and magnetic field vectors, with the strength of the interaction proportional to the change in the axion coupling and the overlap of the waveguide modes at the interface.

There are two possible orientations for the axion coupling, the parallel orientation, where the change in the axion coupling is in the same direction at each interface, $(\Theta_g - \Theta_+) = (\Theta_g - \Theta_-)$, and the anti-parallel orientation, where the change in the axion coupling is in opposite directions each interface, $(\Theta_g - \Theta_+) = -(\Theta_g - \Theta_-)$. Evaluation of the coupling constants shows that in the parallel configuration the TE even mode couples to the TM odd mode and the TE odd mode couples to the TM even mode. In the anti-parallel configuration the TE even mode couples to the TM even mode and the TE odd mode couples to the TM odd mode. In both cases the TE even and odd and TM even and odd modes do not couple and there is no self-coupling. Thus, in each configuration the four coupled mode equations can be split into two independent sets of two coupled mode equations. In the following we focus on the anti-parallel configuration. All other configurations follow identically with the appropriate replacement of mode profiles and coupling constants. In this situation, only the even modes are relevant and the coupled modes equations in Eq. 6 reduce to

$$\frac{d^2}{dx^2}\begin{pmatrix}A_{TE}\\A_{TM}\end{pmatrix} + \begin{pmatrix}\beta_{TE}^2 & \kappa_{TE,TM}\\ \kappa_{TM,TE} & \beta_{TM}^2\end{pmatrix}\begin{pmatrix}A_{TE}\\A_{TM}\end{pmatrix} = 0,$$

(11)

Solving Eq. (11) in the diagonal basis and reverting to the mode basis leads to

$$\begin{pmatrix}A_{TE}(x)\\A_{TM}(x)\end{pmatrix} = \frac{1}{\lambda_0}\begin{pmatrix}(\lambda_+^2 - \beta_{TM}^2)e^{i\lambda_+ x} - (\lambda_-^2 - \beta_{TM}^2)e^{i\lambda_- x} & \kappa_{TE,TM}(e^{i\lambda_+ x} - e^{i\lambda_- x})\\ \kappa_{TM,TE}(e^{i\lambda_+ x} - e^{i\lambda_- x}) & (\lambda_+^2 - \beta_{TE}^2)e^{i\lambda_+ x} - (\lambda_-^2 - \beta_{TE}^2)e^{i\lambda_- x}\end{pmatrix}\begin{pmatrix}A_{TE}(0)\\A_{TM}(0)\end{pmatrix},$$

(12)

with the eigenvalues of the system reading

$$\lambda_\pm^2 = \frac{\beta_{TE}^2 + \beta_{TM}^2 \pm \lambda_0}{2},$$

(13)

with

$$\lambda_0 = \sqrt{(\beta_{TE}^2 + \beta_{TM}^2)^2 + 4\kappa_{TE,TM}\kappa_{TM,TE}}.$$

(14)

One can see that the field amplitude in each mode varies with the sum of two different frequencies. Thus one can find the coupling length and maximum transferred power by considering the beat frequency between the two modes. For a waveguide driven by a plane wave that excites the TE even mode only ($A_{TE}(0) = 1$ and $A_{TM}(0) = 0$), the coupling length, $l_c$, is given by

$$l_c = \frac{\pi}{\lambda_+ - \lambda_-} \approx \frac{\pi\sqrt{2(\beta_{TE}^2 + \beta_{TM}^2)}}{\lambda_0},$$

(15)

where the approximate solution is valid in weak coupling limit, $\kappa_{TE,TM}, \kappa_{TM,TE} \ll \beta_{TE}^2, \beta_{TM}^2$, and which is always valid for the parameters considered here. Similarly, the maximum transferred power is given by

$$P_c = 4\left|\frac{\kappa_{TM,TE}}{\lambda_0}\right|^2.$$

(16)

Given the electric field components, the polarization angle, $\psi$, and ellipticity, $\chi$, are given by

$$\psi = \frac{1}{2}\tan^{-1}\left(\frac{2|E_z|^2}{|E_z|^2 - |E_y|^2}\text{Re}\left[\frac{E_y}{E_z}\right]\right),$$

(17a)

$$\chi = \frac{1}{2}\sin^{-1}\left(\frac{2|E_z|^2}{|E_z|^2 + |E_y|^2}\text{Im}\left[\frac{E_y}{E_z}\right]\right),$$

(17b)

The maximum polarization rotation occurs at the coupling length, $l_c$, and the maximum ellipticity occurs at half the coupling length, $\frac{l_c}{2}$, and can be found from Eqs. (15), (16), (17a) and (17b). In the weak coupling limit these quantities read

$$\psi_{max} \approx \frac{1}{2}\tan^{-1}\left(\frac{2\sqrt{P_c(1-P_c)}}{1-2P_c}\right),$$

(18a)

$$\chi_{max} \approx \frac{1}{2}\sin^{-1}\left(\frac{\sqrt{P_c}(2-P_c)}{2\sqrt{1-P_c}}\right).$$

(18b)